\documentclass[useAMS,usenatbib]{mn2e}
\usepackage{url}
\usepackage{stmaryrd}
\usepackage[pdftex]{graphicx}
\usepackage[draft]{hyperref}
\usepackage{color}
\usepackage[usenames,dvipsnames]{xcolor}

\def\aj{AJ}%
\def\apj{ApJ}%
\def\apjl{ApJ}%
\def\apjs{ApJS}%
\def\aap{A\&A}%
\def\mnras{MNRAS}%
\def\pasp{PASP}%
\def\nat{Nature}%
\def\nSN{1698}

\begin{document}

\title[Matching SN\,Ia Dust Extinction to a host]{The Dependence of the $A_V$ Prior for SN\,Ia on Host Mass and Disk Inclination}

\author[B.W. Holwerda]{B.W. Holwerda$^{1}$\thanks{E-mail: holwerda@strw.leidenuniv.nl, twitter: @benneholwerda}, W.C. Keel$^{2}$, M. A. Kenworthy$^{1}$ and K. J. Mack$^{3,4,5}$\\
$^{1}$ University of Leiden, Sterrenwacht Leiden, Niels Bohrweg 2, NL-2333 CA Leiden, The Netherlands\\
$^{2}$ Department of Physics and Astronomy, University of Alabama, Box 870324, Tuscaloosa, AL 35487\\
$^{3}$ School of Physics (David Caro Building), University of Melbourne, Victoria 3010, Australia\\
$^{4}$ ARC Centre of Excellence for All-sky Astrophysics (CAASTRO), The University of Sydney, NSW 2006, Australia\\
$^{5}$ ARC Centre of Excellence for Particle Physics at Terascale (CoEPP), School of Physics, The University of Melbourne, Victoria 3010, Australia\\
}

\date{Accepted 1988 December 15. Received 1988 December 14; in original form 1988 October 11}

\pagerange{\pageref{firstpage}--\pageref{lastpage}} \pubyear{2010}

\maketitle

\label{firstpage}

\begin{abstract}
Supernovae type Ia (SN\,Ia) are used as ``standard candles" for cosmological distance scales. To fit their light curve shape -- absolute luminosity relation, one needs to assume an intrinsic color and a likelihood of host galaxy extinction or a convolution of these, a color distribution prior. The host galaxy extinction prior is typically assumed to be an exponential drop-off for the current supernova programs ($P(A_V) \propto e^{-A_V/\tau_0}$). We explore the validity of this prior using the distribution of extinction values inferred when two galaxies accidentally overlap (an occulting galaxy pair). We correct the supernova luminosity distances from the SDSS-III Supernova projects (SDSS-SN) by matching the host galaxies to one of three templates from occulting galaxy pairs based on the host galaxy mass and the $A_V$-bias - prior-scale ($\tau_0$) relation from Jha et al. (2007). 

We find that introducing an $A_V$ prior that depends on host mass results in lowered luminosity distances for the SDSS-SN on average but it does not reduce the scatter in individual measurements. This points, in our view, to the need for many more occulting galaxy templates to match to SN\,Ia host galaxies to rule out this possible source of scatter in the SN\,Ia distance measurements. We match occulting galaxy templates based on both mass and projected radius and we find that one should match by stellar mass first with radius as a secondary consideration.

We discuss the caveats of the current approach: the lack of enough radial coverage, the small sample of priors (occulting pairs with HST data), the effect of gravitationally interacting as well as occulting pairs, and whether an exponential distribution is appropriate.
Our aim is to convince the reader that a library of occulting galaxy pairs observed with HST will provide sufficient priors to improve (optical) SN\,Ia measurements to the next required accuracy in Cosmology.

\end{abstract}


\section{\label{s:intro}Introduction}

Supernova Type Ia (SN\,Ia) are a prime Cosmological tool \citep{Riess98,Perlmutter99}. Their relation between light curve and intrinsic luminosity make them excellent standard candles.
Part of the exquisite precision required is a good understanding of the expected dust extinction and known reddening-dimming relation, the Extinction Law \citep[][]{CCM,Calzetti94}. With the Dark Energy detection result in hand, the supernova community now looks for the next step in accuracy \citep[e.g., the 3\% solution][]{Riess11}. A much better model for the dust extinction in the Supernova's host galaxy will be critical. Host galaxy dust is already identified by the Dark Energy Task Force as a principal unknown \citep{DETF} and this remains the case, together with the photometric uncertainty across surveys \citep[e.g.,][]{Conley11, Scolnic14, Betoule14}.

These issues are being addressed by the current generation of supernovae searches, e.g.,  the Supernova Legacy Survey \citep[SNLS,][]{Astier06}, the Equation of State: SupErNovae trace Cosmic Expansion \citep[ESSENCE,][]{Wood-Vasey07}, and the Sloan Digital Sky Survey SuperNova survey \citep[SDSS-SN,][]{Kessler09,Sako14}. The supernovae community has been systematically exploring what could affect the observed relation between peak luminosity and light curve width \citep{Phillips93} for SNe\,Ia. 

A prime candidate is the correlation with properties of the galaxy hosting the SN\,Ia \citep{Gallagher05, Sullivan06, Mannucci06, Pan14a, Kistler13}. Two populations of SN\,Ia appear associated with two populations of stars: blue, star forming galaxies host higher rates of fast declining supernova and red, passive galaxies host predominantly more slowly-declining SNe~Ia \citep{Hamuy96, Howell01, van-den-Bergh05, Mannucci06, Schawinski09, Lampeitl10b, Wang13a}. 
The difference in light curve characteristics may be completely attributable to the different progenitor stellar populations. 

However, simultaneously, the issue of host galaxy extinction has come to the fore, either the applicability of the appropriate extinction law \citep[][]{Riess96b,Phillips99,Altavilla04,Reindl05, Jha07, Conley07}, the validity of the prior of extinction values \citep{Jha07, Wood-Vasey07, Gupta11a}, or the possibility that the geometry of the dusty ISM in (and hence the extinction distribution) may change with galaxy evolution \citep{Holwerda08a,Holwerda15a}. Host galaxies appear to be indistinguishable from normal field galaxies \citep{Childress13}.
After photometric calibration, this host galaxy extinction issue is likely to return as a dominant issue standing in the way of the community's final goal  of 1\% precision for SN\,I \citep[e.g.,][]{Riess11,Kelly14}.

There is now an opportunity to probe the dependencies of the extinction prior through two projects. First, the SDSS-SN project has generated a uniform sample of SN\,Ia and their host galaxy properties of exquisite quality. Secondly, occulting galaxy pairs promise to map extinction in spiral disks using the Hubble Space Telescope ({\em HST}) imaging, enabling a way to generate an extinction prior for SN\,Ia fit. The first of these extinction maps are now available.

In this paper, we explore the effects of applying the extinction priors from occulting galaxy pairs to the SDSS-SN sample and gauge the effects. We opt for the SDSS-SN sample among all the supernova samples available now as it has uniform properties of not only the supernovae, as measured with both light curve packages, but also on the host galaxies. The latter is important if one wants to correct for the effects of host galaxy properties.
The SDSS-SN project uses two light curve fitting packages, {\sc MLCS2k2} and {\sc SALT2}. We use the former's output as it uses an explicit prior for host galaxy extinction, making a direct comparison with the occulting galaxy technique practical, even though SALT and its derivative BaSALT are now often the more preferred package \citep[e.g.,][]{Guy10, Betoule14, Rest14}. 
{ 
We should note here that the $A_V$ value used in {\sc MLCS2k2} is not the exact same quantity as measured from galaxy transparency measurements, i.e., occulting galaxy measurements. {\sc MLCS2k2}'s $A_V$ parameter is the difference in supernova color from an assumed intrinsic one, translated to an extinction via an assumed reddening law \citep[in the case of the SDSS-SN][a grayer one than Milky Way]{Kessler09}. The reddening law in this instant is an implicit input setting, one which has been tweaked to minimize Cosmological fit residuals \citep[e.g.][]{Hicken09}.

We assume in this paper that the shape of the $A_V$ fit parameter's prior and the shape of the distribution of $A_V$ values observed in occulting galaxies is to first order the same.

}

We explore the extinction effects by matching SN\,Ia host galaxies to those occulting galaxies with {\em HST} data through their stellar mass. In a previous paper on SN\,Ia \citep{Holwerda14d}, we explored the relation between the extinction distribution and inclination and found that an inclination correction using a simple cos(i) is sufficient to make the $A_V$ distribution found by the SDSS-SN identical in high and low-inclination subsamples. We therefore adopt this inclination correction in this paper as well. We will discuss the caveats and limits of the current approach and sketch out how to proceed from here in the generation of improved extinction priors, tailored to the host galaxy of the observed supernova.


\section{\label{s:ocg}Occulting Galaxies}

\begin{figure}
  \begin{center}
	\includegraphics[width=0.5\textwidth]{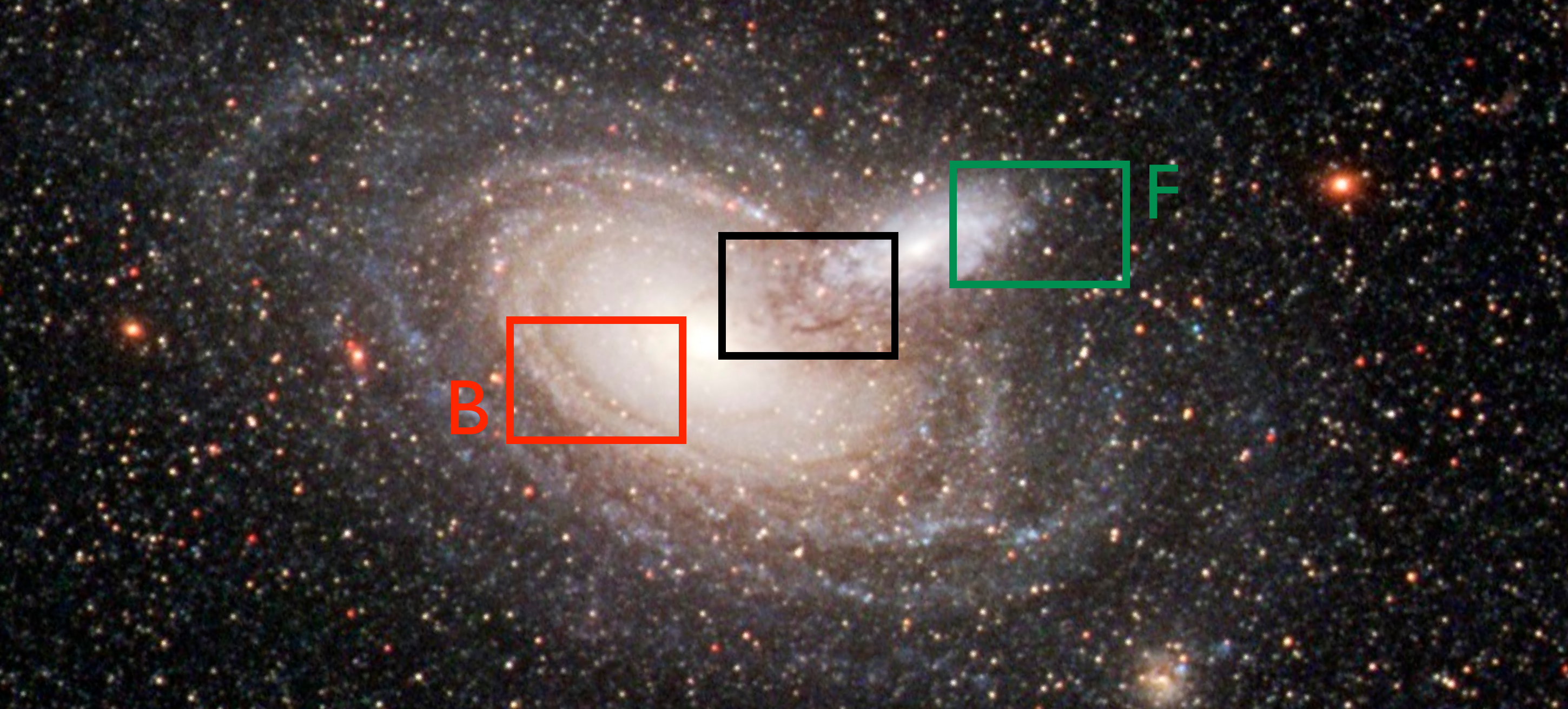}
	\caption{\label{f:pair}The 2MASX occulting galaxy pair at z=0.06. The extinction in the overlap region ({\bf black aperture}) can be estimated from the complementary apertures; the foreground spiral ({\color{ForestGreen}F, the green aperture}) and the background galaxy ({\color{red}B, red aperture}).}
  \end{center}
\end{figure}

One can accurately measure the extinction of light by interstellar dust using a {\em known} light source. The most accurate extra-galactic method is to use a partially occulting galaxy pair (Figure \ref{f:pair}). Assuming both galaxies are symmetric, one can estimate flux contributions to the overlap region from the complementary regions at the same radius in each galaxy. 
The reliance on differential photometry enables an extinction measurement ($A_V$) in a single-color image without relying on any assumed Extinction Law. 
With multiple filters, the effective Extinction Law itself can be measured \citep{kw01a,Holwerda09,Keel14}. 

\subsection{The Occulting Galaxies Method}

Figure \ref{f:pair} shows an example pair, serendipitously imaged with {\em HST} for the ANGST survey \citep{angst}, near NGC 253 (2MASX) To determine the extinction in the overlap region ({\bf black aperture}), one estimates the relative flux contributions by both galaxies from the complementary apertures in the foreground ({\color{ForestGreen}green aperture}) and background galaxy ({\color{red}red aperture} in Figure \ref{f:pair}), respectively, assuming rotational symmetry (i.e. the image be can rotated and be self-similar). Thus, we estimate the flux contributions to the overlap region from the same image. In the case of ground-based images, each of these apertures would yield single $A_V$ value, averaged over a large portion of the foreground disk. In the case of an {\em HST} image, it results in a highly accurate {map} of the dust extinction in the overlap region with hundreds of independent lines-of-sight. 
The map is constructed pixel-by-pixel; for each pixel in the overlap aperture, the corresponding pixel in the background galaxy aperture (red) is found i.e., the pixel at the same distance from the background galaxy center but in the opposite direction, as well as the corresponding pixel in the foreground galaxy (same distance with respect to the foreground galaxy center).
We estimate the optical depth ($\bar{\tau}$) or extinction ($A_V$) from the flux in the overlap ($F+Be^{-\tau}$), a mix of flux from background and foreground galaxies, the flux in the corresponding background pixel ({\color{red}$\bar{B}$}) and foreground pixel ({\color{ForestGreen}$\bar{F}$}, same colors as Figure \ref{f:pair}):
\begin{equation}
A_V = 1.086 \times \bar{\tau} = - 1.086 \times ln \left( {(F+Be^{-\tau}) - {\color{ForestGreen} \bar{F}} \over {\color{red} \bar{B}}} \right)
\end{equation}

Thus, a map of dust extinction and a distribution function ($P(A_V)$) such as Figure \ref{f:pav} can be constructed from an {\em HST} image. For each type of foreground galaxy, inclination and radial distance, an appropriate $P(A_V)$ can thus be observed.

\subsection{Occulting Pairs}

\begin{figure}
\begin{center}
\includegraphics[width=0.5\textwidth]{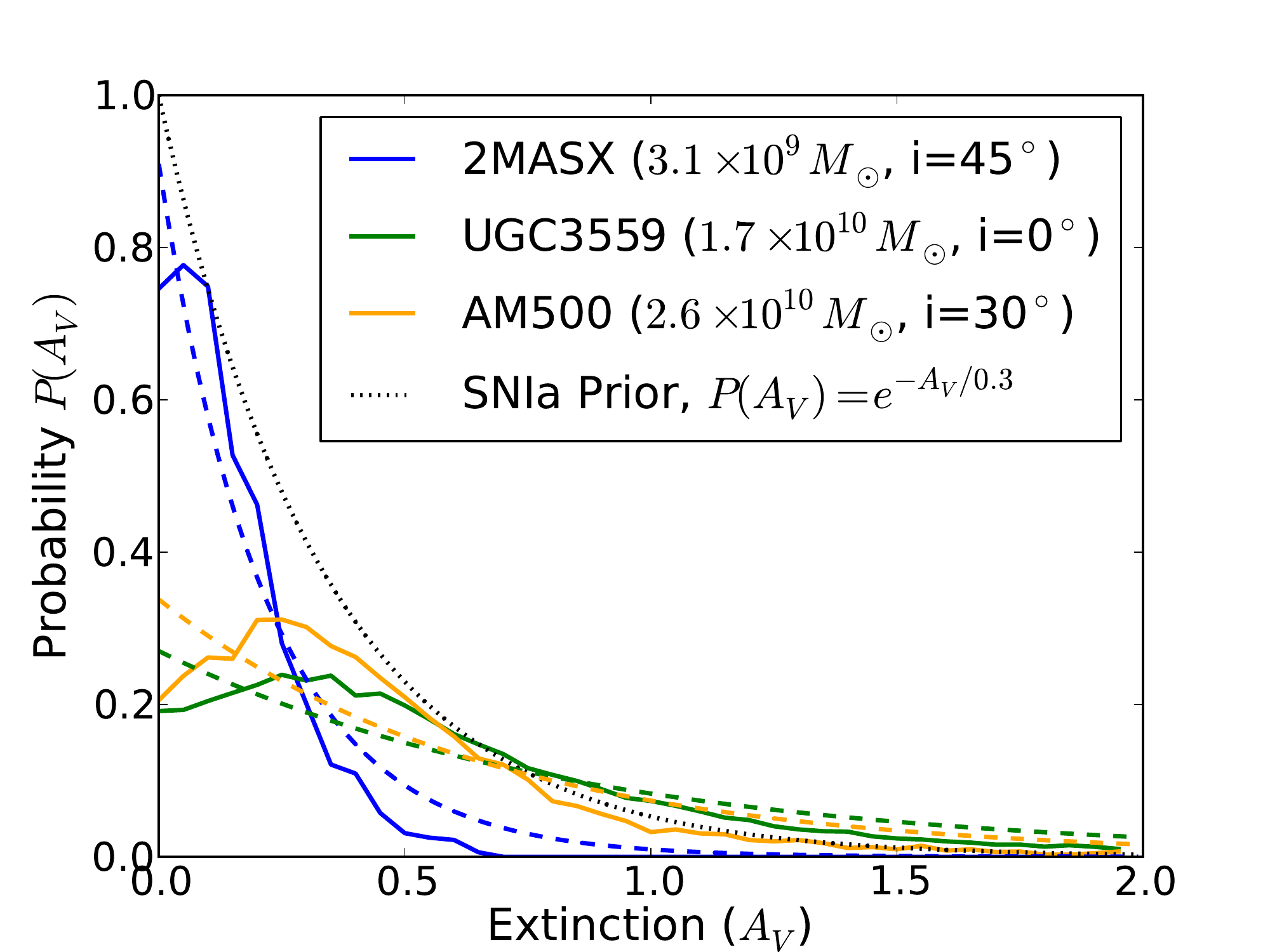}
\caption{\label{f:pav} The shape of the $A_V$ distribution of the three occulting galaxies with enough overlap to generate a distribution. These distributions have not yet been corrected for their foreground disk's inclination. Dashed lines are exponential fits through the distributions. The exponential drop-off can then be used as a SN\,Ia prior. The derived drop-offs are listed in Figure \ref{f:Avbias}, with their $\tau_0$ values listed in Table \ref{t:ocg}.}
\end{center}
\end{figure}

The accuracy in individual pairs is limited by the symmetry of both galaxies and image quality. The uncertainty from the assumption of galaxy symmetry can mitigated by using elliptical galaxies as the background galaxy and by combining the results for several pairs. In the case of {\em HST} images (e.g., Figure \ref{f:pair}), a single pair contains enough independent lines-of-sight through the foreground galaxy to construct a $P(A_V)$ such as Figure \ref{f:pav}.

The occulting galaxies technique however has increased steadily in accuracy and usefulness, owing in a large part to the increasing sample sizes. Estimating dust extinction and mass from differential photometry in occulting pairs of galaxies was first proposed by \cite{kw92}. Their technique was then applied to all known pairs using ground-based optical images \citep{Andredakis92, Berlind97, kw99a, kw00a} and spectroscopy \citep{kw00b}, and later space-based {\em HST} images \citep[][]{kw01a, kw01b, Elmegreen01,Holwerda09, Holwerda12}. These initial results were limited by sample sizes. More recently, more pairs were found in the SDSS spectroscopic catalog \citep[86 pairs in][]{Holwerda07c}, and in an ongoing effort in the GalaxyZOO project \citep{galaxyzoo}. The current count  stands at 1993 pairs \citep{Keel13} providing opportunities for follow-up with IFU observations \citep{Holwerda13a,Holwerda13b} and GALEX \citep{Keel14}. 

Highly complete spectroscopic surveys, such the Galaxy and Mass Assembly 
 \citep[GAMA][]{Driver09, Driver11, Baldry10} are adding many more overlapping pairs through the identification of blended spectra or close on-sky pairs (Holwerda et al. {\em in prep.}).
Observational results include mean extinction profiles \citep{kw00a,kw00b,Holwerda07c}, an indication that the dust may be fractal \citep{kw01a} and that the color-extinction relation is gray in the ground-based observations \cite{Holwerda07c}. The Galactic Extinction Law returns as soon as the physical sampling of the overlap region resolves the molecular clouds in the foreground disk \citep[$<100$pc,][]{kw01a,kw01b, Elmegreen01,Holwerda09}.

The extinction maps of foreground disks made with {\em HST} are especially useful as a probability map of extinction towards a single object of interest. To date, we have obtained {\em HST} imaging of three overlapping pairs with a large enough overlap region to generate a reasonable distribution. The normalized distributions of these three are shown in Figure \ref{f:pav}. Radial coverage is very similar to the radii at which SN\,Ia are found (Table \ref{t:ocg}): between 0 and 6 effective radii \citep[converted using the prescriptions in ][]{Graham05}. 

Our previous results on occulting galaxy pairs are consistent with an exponential decline with radius of the average disk attenuation, not dissimilar to the exponential profile of the stellar light \citep[e.g.,][]{kw99,Holwerda05, Holwerda05a, Holwerda05b}. We stress here that we deal here now in the exponential distribution of values in a certain range of radii but this is not the same as the radial exponential decline. For example, the lower values ($A_V\sim0.1$) are typically, but not exclusively, found at higher radius while the higher values are more typical for spiral arms.



\begin{table}
\caption{The properties of the foreground occulting galaxy in three pairs with sufficient {\em HST} information to form an extinction map. Stellar mass in solar masses and radii in effective (half-light) radii.}
\begin{center}
\begin{tabular}{l l l l l}
Pair			& Type 	& Stellar Mass			& Radial Coverage	& \\
			& 		& $M^*$				& ($R_e$)			& ($\tau_0$)\\
\hline
UGC 3995	& Sa		& $1.7 \times 10^{10}$	& 0 -- 2			& 0.22 \\ 
2MASX		& Sd		& $3.1 \times 10^9$		& 2 -- 5			& 0.85 \\
AM 0500		& Sbc	& $2.6 \times 10^{10}$	& $\sim 3.7$		& 0.67 \\
\hline
\end{tabular}
\end{center}
\label{t:ocg}
\end{table}%

\subsection{UGC 3995}

This pair of galaxies has been long known to be overlapping and interacting \citep[e.g.,][]{Marziani99} and we have analyzed the extinction properties of both CALIFA DR1 IFU data and archival WFPC2 {\em F606W} imaging in \cite{Holwerda13b}. 
Radial coverage is from 5-15", covering the full 15\farcs60 Petrosian radius (SDSS-r) of UGC3995B, the foreground galaxy.
We obtained the foreground galaxy's mass from the SDSS fluxes of the unocculting side and the prescription from \cite{Zibetti09a}.

\subsection{2MASX}

This pair's serendipitous discovery was originally reported in \cite{Holwerda09}. The distance and  stellar mass was estimated in \cite{Holwerda13a} and the radial coverage of the overlap region is between 2 and 5 effective (or half-light) radii ($R_e$).

\subsection{AM 500--620}

This was a pair identified in \cite{kw00a} as an overlapping galaxy pair and the {\em HST} data were originally reported in \cite{kw01a}. The radial coverage is mostly in the outer disk \citep[$\sim R_{25}$, the 25 mag/arcsec$^2$ as defined by][]{RC3}. Stellar mass and Hubble type are from  \cite{kw01a}. 


\begin{figure}
\begin{center}
\includegraphics[width=0.5\textwidth]{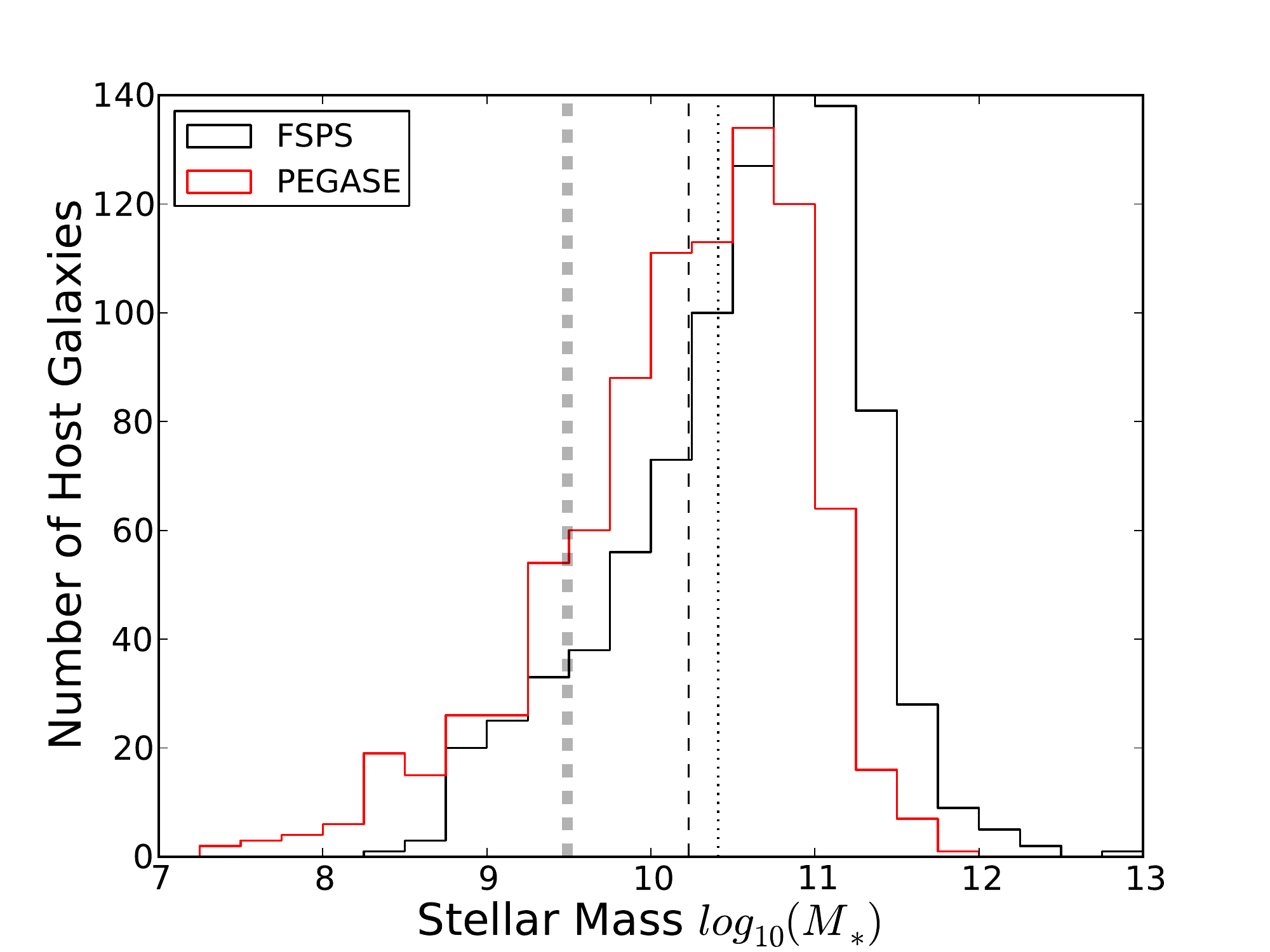}
\caption{The distribution of SN\,Ia host galaxy mass from \protect\cite{Sako14}. They present both FSPS \citep{FSPS} and PEGASE \citep{Bruzual09} stellar masses. The masses of 2MASX (thick dashed), UGC 3995 (dashed line), and AM500 (dotted line) are marked as well.}
\label{f:mass}
\end{center}
\end{figure}

\section{\label{s:sn}SuperNovae Type Ia Sample}

Type \,Ia Supernovae (SN\,Ia) are the touchstone of Cosmological distance measurements. 
The photometric properties of their light curves are very stable but appear to depend slightly 
on host galaxy properties (e.g., star-formation rate or stellar mass).
With this mind, the third incarnation of the Sloan Digital Sky Survey (SDSS-III) included a 
supernova search with spectroscopic follow-up. This SDSS-Supernova search (SDSS-SN)
\footnote{\url{http://www.sdss.org/supernova/aboutsupernova.html}}  has yielded a wealth of galaxy and supernova properties.

SDSS-SN is described in \cite{Frieman08} and in more detail in \cite{Sako08} and \cite{Sako14}. 
Observing for three months of the year from 2005 to 2007, SDSS-SN identified hundreds
spectroscopically confirmed SNe \,Ia in the redshift range 0.05 $<$ z $<$ 0.35. 

\cite{Sako14} present the final data-product of this tremendous observational effort. They include 
an assessment of SN type based on the light curve (PSNID output) and light curve fits using the 
two most commonly used packages, {\sc SALT2} \citep{Guy07} and  {\sc MLCS2k2} \citep{Jha07}. 
Of these two packages, the {\sc SALT2} light curve fitter is now the most popular. Yet we focus on 
the {\sc MLCS2k2} package because it explicitly starts with a host galaxy extinction prior, making 
it easier to directly compare its assumptions to our observations in occulting pairs.

{ 
However, we should note that both SALT2 and {\sc MLCS2k2} essentially measure the reddening
of the supernova, not actual attenuation but the convolution of intrinsic supernova color distribution,
scatter by the surrounding interstellar matter and dust attenuation in the host galaxy.
}
The shape of the {\sc MLCS2k2} $A_V$ prior used for the light curve fits was: $P(A_V) \sim e^{-A_V/\tau_0}$, with $\tau_0 =0.4$. Because it has host galaxy extinction as an explicit prior and a model relation between $A_V$ distribution width ($\tau_0$) and $A_V$ bias, we use the {\sc MLCS2k2} values for our further analysis here. 
{ 
The {\sc MLCS2k2} version used for the SDSS-SN analysis assumed a $R_V=2.8$ reddening relation to convert reddening to extinction $A_V$ \citep{Kessler09}, indicating it is not completely host attenuation alone. However, it is not as unphysical as the $R_V=1.7$ needed elsewhere with {\sc MLCS2k2} \citep{Hicken09}. 
For our purposes here, we will explicitly assume that the shape of the prior for  {\sc MLCS2k2}'s fit parameter $A_V$ can be taken to be similar to the distribution of actual attenuation values as found in occulting galaxies.
}

Galaxy properties are those available from the SDSS-DR9 database and stellar mass and star-formation are modeled with two different packages, FSPS \citep{FSPS} and PEGASE \citep{Bruzual09}. For the purpose of this paper, the stellar mass is of interest and we choose the FSPS value for for the further analysis. Figure \ref{f:mass} shows the distribution of host mass galaxies with SN\,Ia. The mean mass is $\sim10^{10} M_\odot$, similar to AM500. 
We focus only on those objects that have reasonable chance of being bona-fide SN\,Ia (S/N$>$5 and P(SN\,Ia) $>$ 50\%) to ensure the conclusions for the improved prior are based on these only (\nSN\ SN in the total sample).

\begin{figure}
\begin{center}
\includegraphics[width=0.5\textwidth]{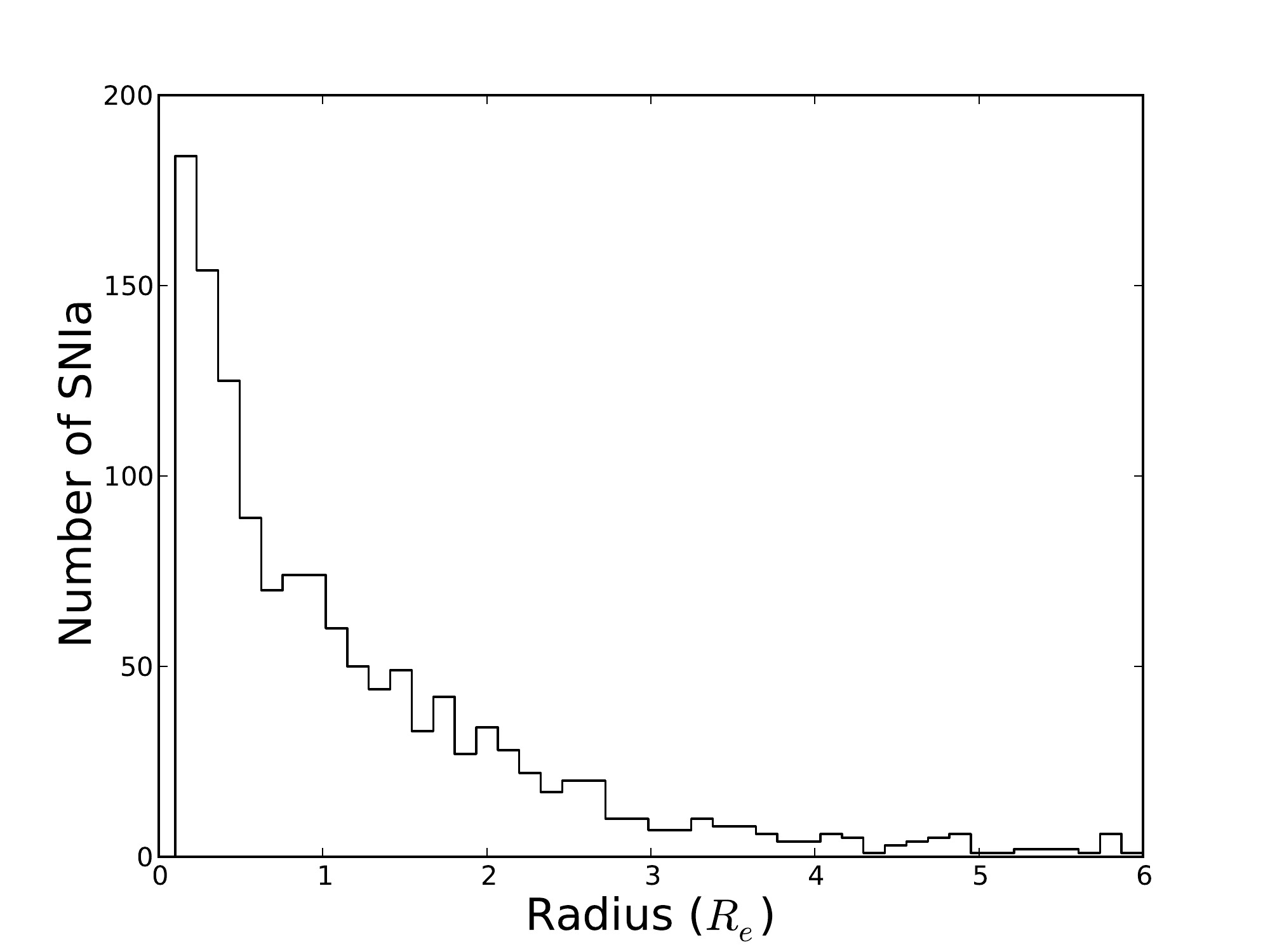}
\caption{The distribution of the projected separation of  SN\,Ia from their host galaxy in effective radii ($R_e$) based on the values reported in \protect\cite{Sako14}. For comparison,  \protect\cite{Sako14} plot the distance-to-host  or "directional light radius" ($d_{DLR}$), which is not deprojected into the plane of the host disk.}
\label{f:Re}
\end{center}
\end{figure}

In addition to the values thoughtfully provided by the SDSS-SN project in \cite{Sako14}, we retrieve the axis ratio, position angle, and Petrosian radius from the SDSS server. Using these values, we compute the galactocentric radial distance from the center of the host galaxy in Petrosian radii. This is a different value than that of the separation presented in \cite{Sako14}, the "directional light radius" ($d_{DLR}$), which is not deprojected into the plane of the host disk.
If we assume disks (n=1 S\'{e}rsic index), the Petrosian radius corresponds to approximately two effective radii \citep[$R_p = 2.15 \times R_e$, see][]{Graham05}. Figure \ref{f:Re} shows the radial coverage for the SN\,Ia for which SDSS Petrosian radii were available. SN\,Ia are typically found by SDSS-SN between 0 and 5 effective radii ($< 2.5$ Petrosian radii), which correspond approximately to the radii covered by the occulting galaxy pairs (Table \ref{t:ocg}).

\section{\label{s:analysis}Analysis}

\begin{figure}
\begin{center}
\includegraphics[width=0.5\textwidth]{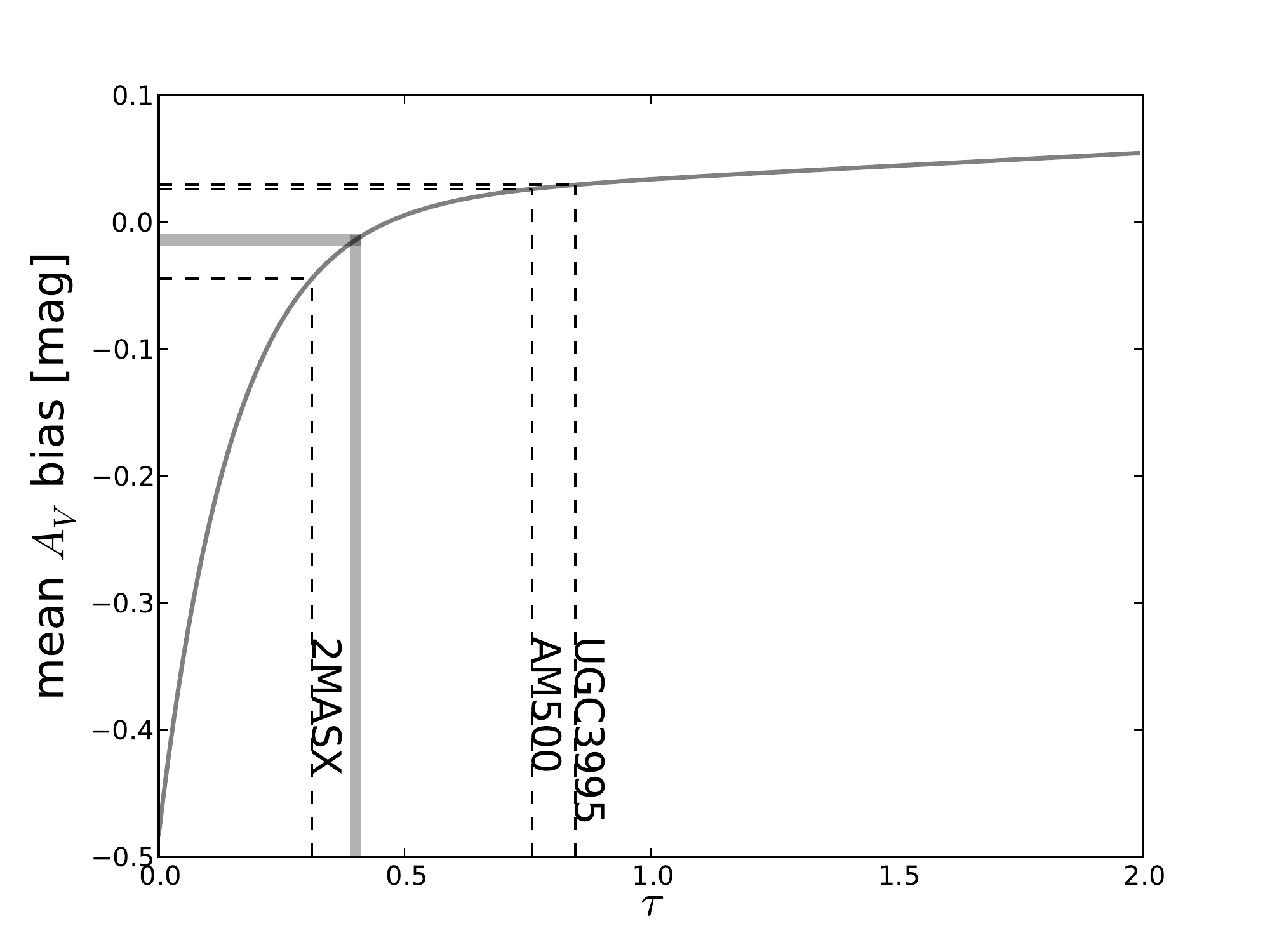}
\caption{The relation between prior exponential scale and the resulting bias in the $A_V$ from \protect\cite{Jha07}. The thick gray line is the MLCS2K2 prior. The values of the three occulting pairs are also marked (not corrected for their inclination).}
\label{f:Avbias}
\end{center}
\end{figure}

\subsection{$A_V$ Distribution and Bias}

The relation between the exponential drop-off and bias introduced in the $A_V$ value was explored by \cite{Jha07} using the \cite{Riello05} simulations. Figure \ref{f:Avbias} shows their relation \citep[Figure 22 in the appendix of][]{Jha07}. We mark the $\tau_0$ values inferred from the occulting galaxy pairs, corrected for inclination ($\tau_0 = \tau/cos(i)$. 
To parameterize the $A_V$ bias, we fit the relation in Figure \ref{f:Avbias} with:
\begin{equation}
\label{eq:Avbias}
\Delta A = -0.5 \times e^{-\tau_0/0.15}  +0.02 \times \tau_0,  
\end{equation}
\noindent where $\tau_0$ is the exponential drop-off of the host galaxy prior under consideration.

We note first that none of our three occulting galaxy templates match the drop-off typically used in MLCS2K2 ($\tau_0 = 0.4$) which was used for the SDSS-SN analysis \citep{Sako08,Sako14}. Therefore, assuming the SDSS-SN galaxies adhere to one of these three templates based on their host mass will inevitably introduce a bias in the distribution. Our goal is to explore in which direction and how severe it would be.

\begin{figure}
\begin{center}
\includegraphics[width=0.5\textwidth]{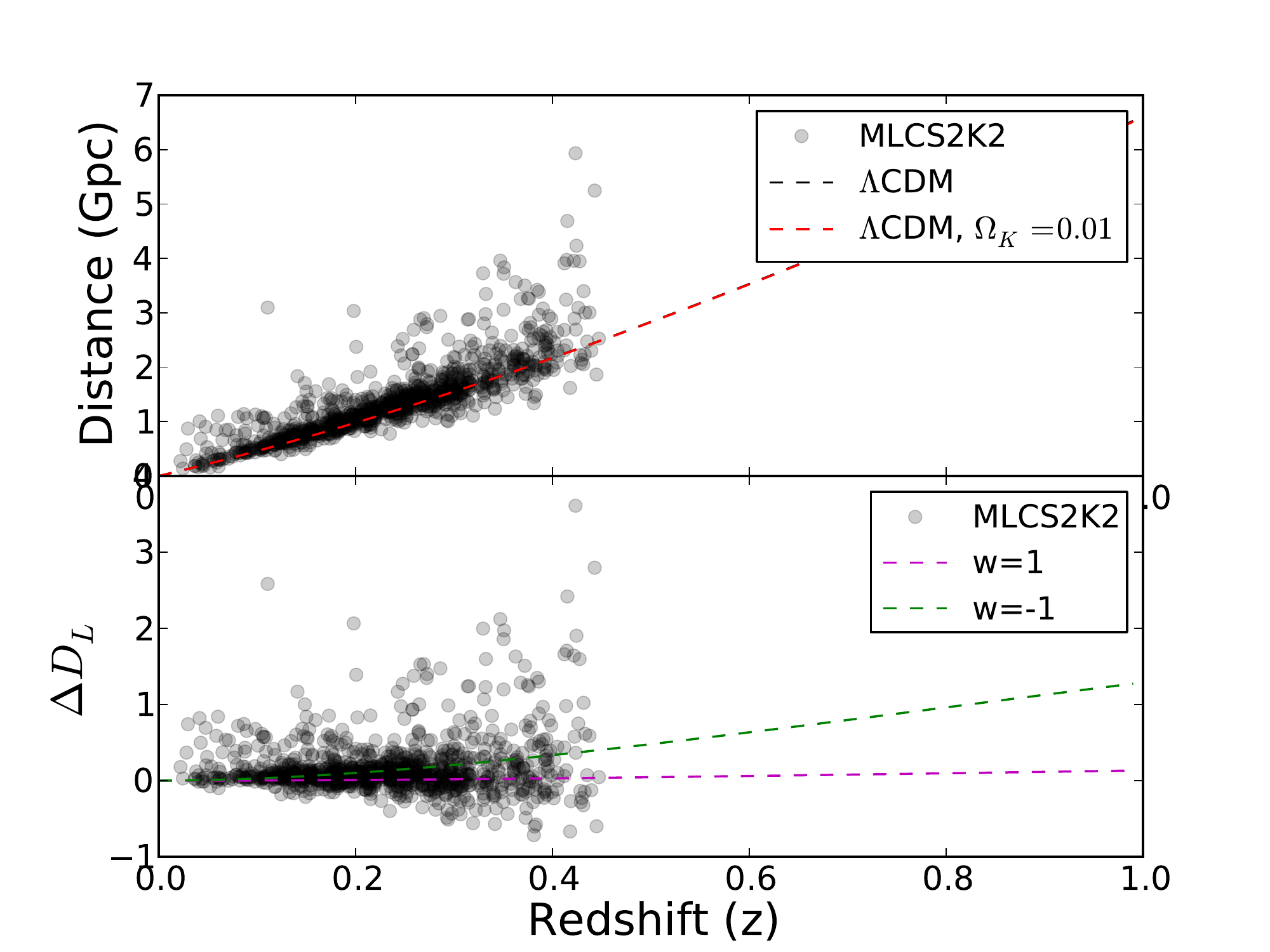}
\caption{The relation between redshift and MLCS2K2 luminosity distance (top panel) and the difference between the $\Lambda$CDM and the SN\,Ia luminosity distances (bottom panel). The relation for different equations of state ($w=1/-1$) are shown for illustration only.}
\label{f:zD}
\end{center}
\end{figure}

\subsection{New Luminosity Distances for the SDSS-SN}

Figure \ref{f:zD} shows the relation between the luminosity distance and redshift for the SDSS-SN Sn\,Ia.
They closely follow the expected $\Lambda$CDM relation with scatter around the mean in the residual (bottom panel in Figure \ref{f:zD}). To extract the exact nature of Dark Energy, one approach is to explore this residual. However, inconsistent or heterogeneous photometry or host galaxy extinction could introduce biases in this residual.

\begin{figure}
\begin{center}
\includegraphics[width=0.5\textwidth]{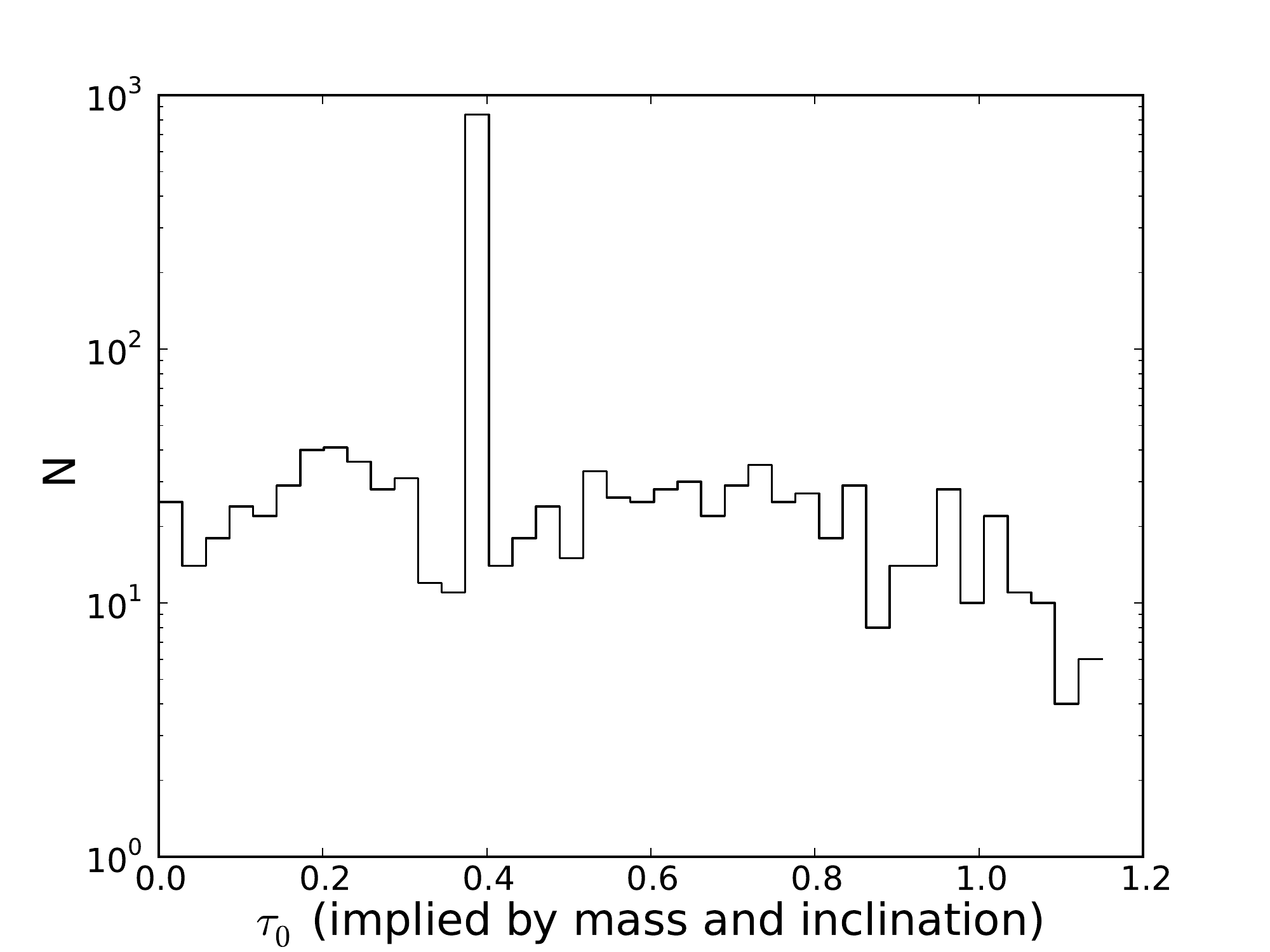}
\caption{The distribution of host galaxy $\tau_{host}$ values as calculated from equation \ref{eq:realtau}. There is still a peak at $\tau_0=0.35$ as this is the value assigned when no good host estimate was available. The concentration at $\tau_0=0.$ is due to poorly constrained inclination. }
\label{f:realtau}
\end{center}
\end{figure}

We assign each host galaxy a new exponential template ($\tau_{OCG}$) from one of the three occulting galaxy (OCG) distributions (fit as an exponential drop-off), based on the nearest in stellar mass, and correct for disk inclination:

\begin{equation}
\label{eq:realtau}
\tau_{host} = \tau_{OCG} {\rm (M)} \times cos(i_{host})
\end{equation}
\noindent where $i_{host}$ is the inclination of the host galaxy as inferred from the minor and major axes ratio: 
\begin{equation}
\label{eq:incl}
cos(i_{host}) = \sqrt{ {(b/a)^2 - (b/a)_{min}^2 \over 1- (b/a)_{min}^2}},
\end{equation}
\noindent originally from \cite{Hubble26}, where $a$ is the major axis, $b$ the minor axis and we assume $b/a_{min} =0.1$.
We found in Holwerda et al. ({\em submitted}) that the distribution of SDSS-SN $A_V$ values becomes the same for high- and low-inclination galaxies, once the values are corrected for disk inclination using this simple $cos(i)$ correction (i.e., the screen approximation). 

Figure \ref{f:realtau} shows the distribution of resulting $\tau_{host}$ for the SDSS-SN SN\,Ia sample. The peak around $\tau_{host}=0.4$ corresponds to all the host galaxies for which no mass estimate was available in the SDSS-SN catalog.

\begin{figure}
\begin{center}
\includegraphics[width=0.5\textwidth]{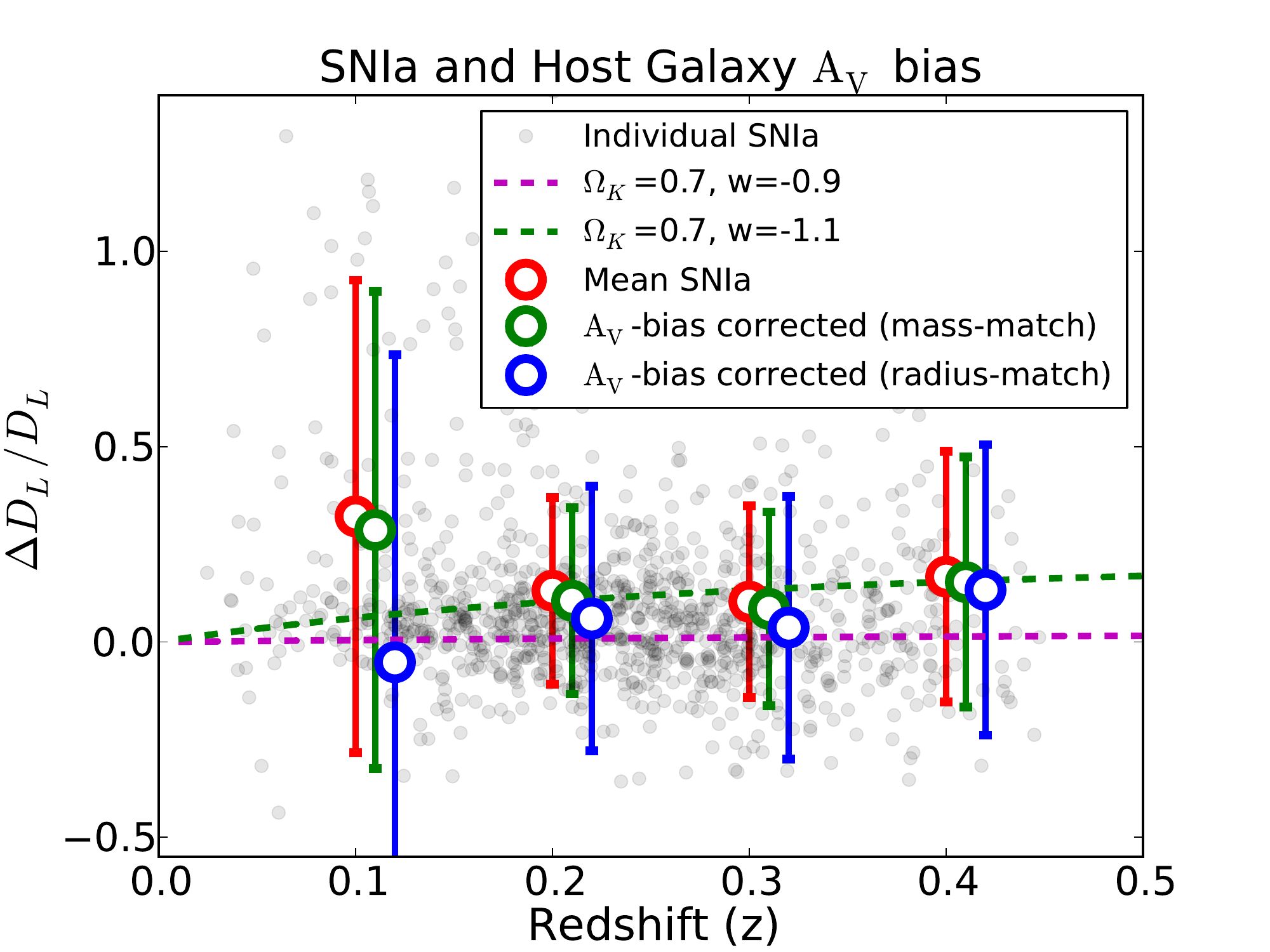}
\caption{The relative difference between $\Lambda$CDM predicted distance and the SN\,Ia luminosity distances from SDSS-SN. Individual SN\,Ia (grey points) and the mean and rms (red points) and once corrected for $A_V$ bias (green points) using the occulting galaxy templates (Figure \ref{f:pav}), matched through host mass. Alternatively, one can match OCG templates by radius (blue points). Two Cosmologies are shown as well to illustrate: w=-0.9 and -1, neither a fit to the data. The different Cosmologies are shown to illustrate how the choice of dust prior can have an almost similar effect on the inferred Cosmology.}
\label{f:zDeltaL}
\end{center}
\end{figure}

Starting from the new $\tau_{host}$ exponential values for each host galaxy (equation \ref{eq:realtau}), we can now calculate the $A_V$ bias for each host, based on its stellar mass and inclination (equation \ref{eq:Avbias}) and apply this to the luminosity distance.
Figure \ref{f:zDeltaL} shows the relative difference with $\Lambda$CDM distances as a function of redshift and the mean values (and dispersion) before our $A_V$ bias correction and after (Table \ref{t:zDL}). 

There are two things to note in the corrected values. First is that the mean luminosity distance is slightly lower, i.e., there is a little more extinction in the SN\,Ia host galaxies if we use individual priors, but there is no less scatter in the values. 

The change in the mean is interesting for cosmological study but at present only points to the direction one can expect the luminosity distances to change. The priors may be slightly more tailored but they are hardly appropriate for each host galaxy (see below for the occulting galaxy pair prior caveats). 
This lack of an appropriate prior is reflected --in our opinion-- in the lack of change in dispersion. 
If an appropriate correction to the prior shape and hence $A_V$ bias had been applied, luminosity distances would adhere closer to the Hubble flow and any remaining scatter can be attributed to individual motion of the galaxies or photometric errors.

Alternatively, one could match the few occulting galaxy templates by their projected radius, not the foreground galaxy's mass. These are the blue averages points in Figure \ref{f:zDeltaL}. The scatter around the new mean residuals is worse than in the case of the mass-matched templates. We can speculate at this point that to match templates, one should select by host mass first and galactocentric radius only as a secondary effect. Once again, we are restricted by the numbers of occulting pair templates we have.

%
%
%
%

\begin{table}
\caption{The average and scatter for $\Delta_L/D_L$ in four bins for the original MLCS2K2 analysis, one corrected using the occulting galaxy templates matched with host mass or radius.}
\begin{center}
\begin{tabular}{l l l l}
z	& $\Delta_L/D_L$ & Mass-Matched	& Radius Matched \\
\hline
\hline
0.10 & 0.32 (0.60)      & 0.29 (0.61)          & -0.05 (0.79) \\
0.20 & 0.13 (0.24)      & 0.11 (0.24)          & 0.06 (0.34) \\
0.30 & 0.10 (0.25)      & 0.08 (0.25)          & 0.04 (0.34) \\
0.40 & 0.17 (0.32)      & 0.15 (0.32)          & 0.13 (0.37) \\
\hline
\end{tabular}
\end{center}
\label{t:zDL}
\end{table}%

\section{Caveats}

While the radial coverage and host mass range appears reasonably approximated by our three template host galaxies, there are a number of issues outstanding with just using three galaxies as templates. We will focus on the shortcomings of the occulting galaxy pairs presented as priors.

\subsection{MLCS2K2 $A_V$ parameter is all Dust Attenuation}

The central assumption in this work has been that the attenuation seen in occulting galaxies is the same as the $A_V$ parameter in MLCS2K2. 
However, because the $A_V$ is derived assuming an attenuation law, it cannot be ruled out that some measure of the reddening is not from dust attenuation but scatter in, or SN intrinsic color scatter.

\subsection{Radial Coverage}

While Table \ref{t:ocg} and Figure \ref{f:Re} seem to suggest that at least the SN\,Ia occur mostly at the same radii for which we have some extinction distribution from occulting pairs, the coverage is far from complete. For example, for low-mass galaxies, there is only information between 1-2 effective radii. 

\subsection{Small Sample}

It is obvious to point out that it is impossible to infer a relation between stellar mass and an extinction distribution with only three galaxy templates. We plan to obtain more templates from future {\em HST} observations \citep{Holwerda14b} but for now we can only assign the exponential prior to the closest stellar mass to the host mass. Alternatively, we fitted a linear relation between $\tau_0$ and mass for the occulting galaxy pairs but this resulted in no improvement of the variance of the SN\,Ia distances.

\subsection{Interactions}

Occulting galaxies are often interacting and the redshift difference for two of our templates (UGC 3995 and 2MASX) are such that an ongoing interaction cannot be ruled out. It is very likely in the case of UGC 3995. The exact effect of gravitational interaction on the distribution of ISM is unknown but likely to be severe. Is there more dust now at higher radii? Do the shocks and tides move more dust into denser clouds in advance of the burst of star-formation associated with gas-rich interactions? The way around this issue again is to observe more and bona-fide occulting pairs, well separated in redshift. A selection from a spectroscopic survey such as the Galaxy And Mass Assembly Survey \citep[GAMA][]{Driver09, Driver11, Baldry10} may well prove to be ideal.

\begin{table}
\caption{The exponentially modified Gaussian fits to the $A_V$ distributions observed in occulting pairs.}
\begin{center}
\begin{tabular}{l l l l}
pair	& m & $\sigma$ &  $\lambda$ \\
\hline
\hline
2MASX & 0.013 &  0.015 &  6.65 \\
UGC3559 & 0.092 &  0.092 &  2.10 \\
AM500 & 0.013 &  0.011 &   2.29 \\
AM1316 &  1.41 &  0.50 &   70.60 \\
\hline
\end{tabular}
\end{center}
\label{t:expGaus}
\end{table}%

\begin{figure}
\begin{center}
\includegraphics[width=0.5\textwidth]{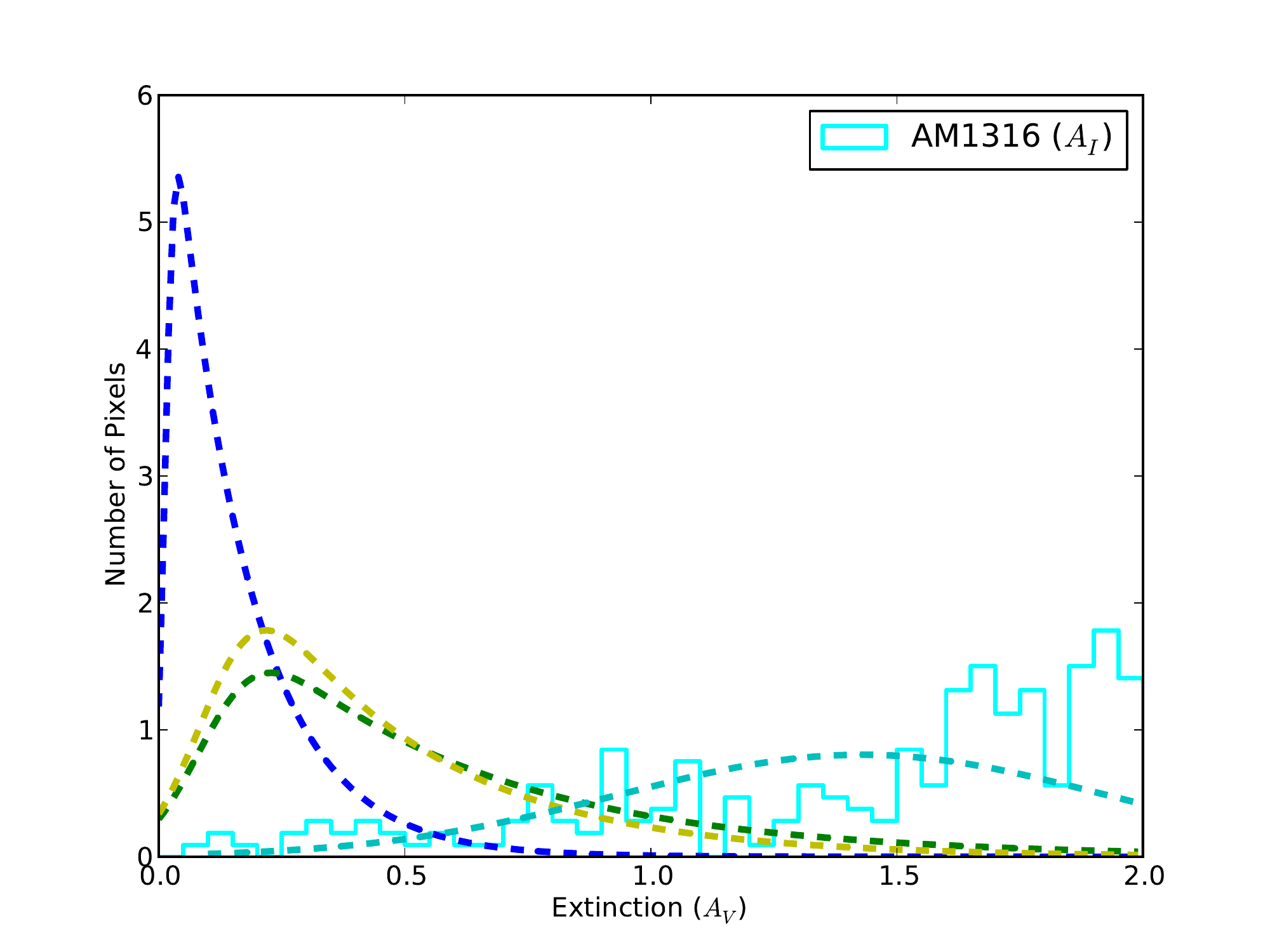}
\caption{\label{f:mle} The shape of the $A_V$ distribution of the three occulting galaxies with enough overlap to generate a distribution. An exponentially modified Gaussian distribution was with using maximum likelihood to each distribution. The relevant parameters are listed in Table \ref{t:expGaus}.}
\end{center}
\end{figure}

\subsection{Is the $A_V$ prior really an exponential?}

One strong assumption so far has been that the distribution of extinction values is an exponential drop-off. However, Figure \ref{f:pav} shows that none of the three distributions is fully described as an exponential. For example, in the more massive galaxies, the peak of the distribution is not close to $A_V=0$ at all. We noted this in \cite{Holwerda13b} for both the {\em HST} observations and the IFU ones. In the case of UGC 3995, one could contribute this to the ongoing interaction. Again the only way forward it to obtain more templates for the same host mass and infer the mean and variance of the distribution function. 

Our assumption --and the MLCS2K2 authors'-- is that the shape of the distribution is an exponential. We now strongly suspect the true shape of the extinction distributions may be a log-normal one, similar to what Dalcanton et al. ({\em in preparation}) find for the extinction values in front of stars in M31 based on the PHAT project \citep{PHAT} or an exponentially modified Gaussian. To illustrate, Figure \ref{f:mle} shows fits to the $A_V$ distribution using the exponentially modified Gaussian fit to each distribution using maximum likelihood. 

This strongly suggests that MLCS2K2 will now have to be redone using these templates instead of an exponential one.

%
%

\begin{figure}
\begin{center}
\includegraphics[width=0.5\textwidth]{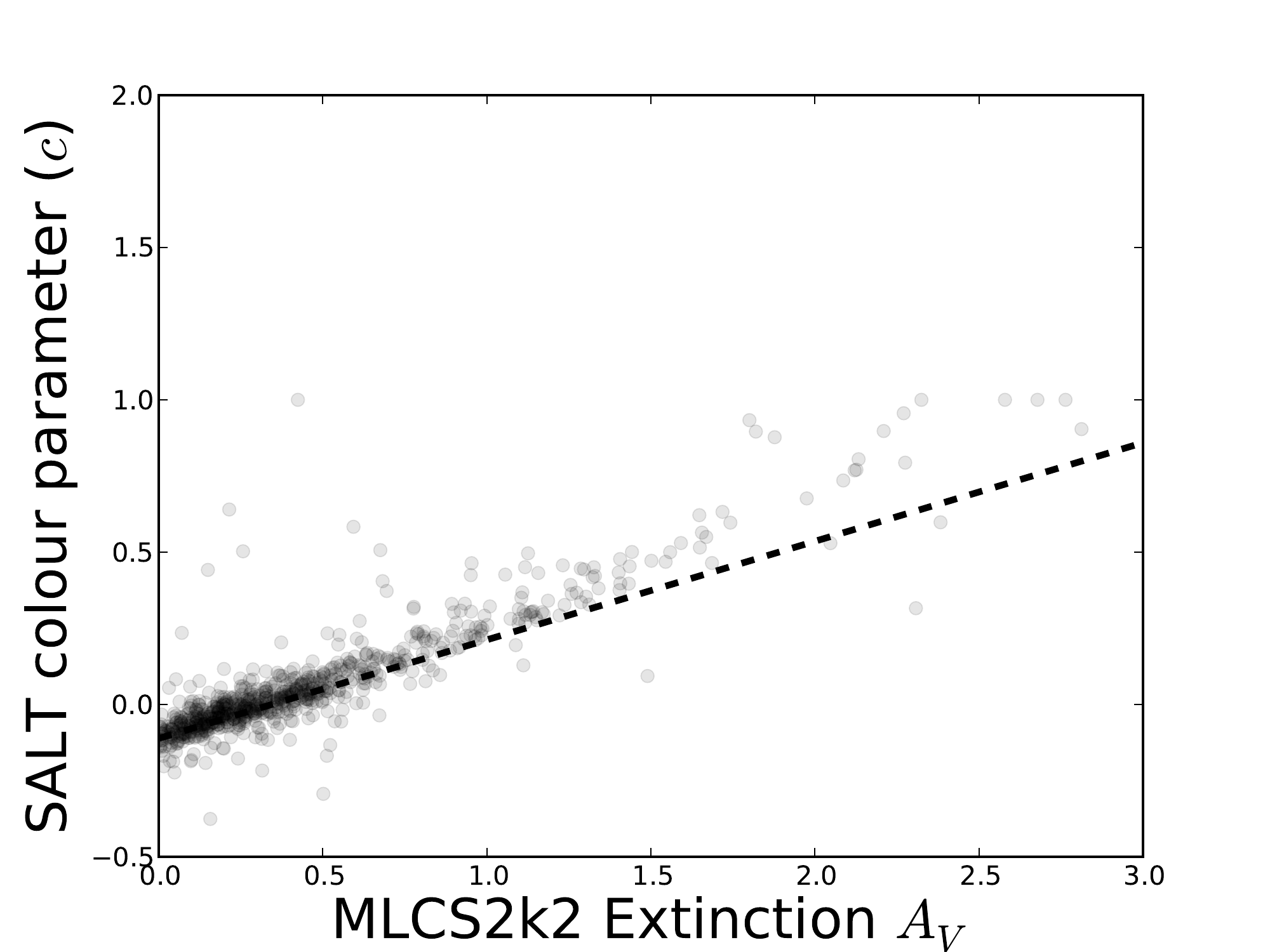}
\caption{The relation between extinction as measured by the {\sc MLCS2k2} package and the color parameter from the {\sc SALT2} package for the SDSS-SN supernovae. There is a linear relation between the two (fit with maximum likelihood, including the {\sc SALT2} color errors.}
\label{f:salt:c}
\end{center}
\end{figure}

\section{Application to {\sc SALT2} }

\begin{figure}
\begin{center}
\includegraphics[width=0.5\textwidth]{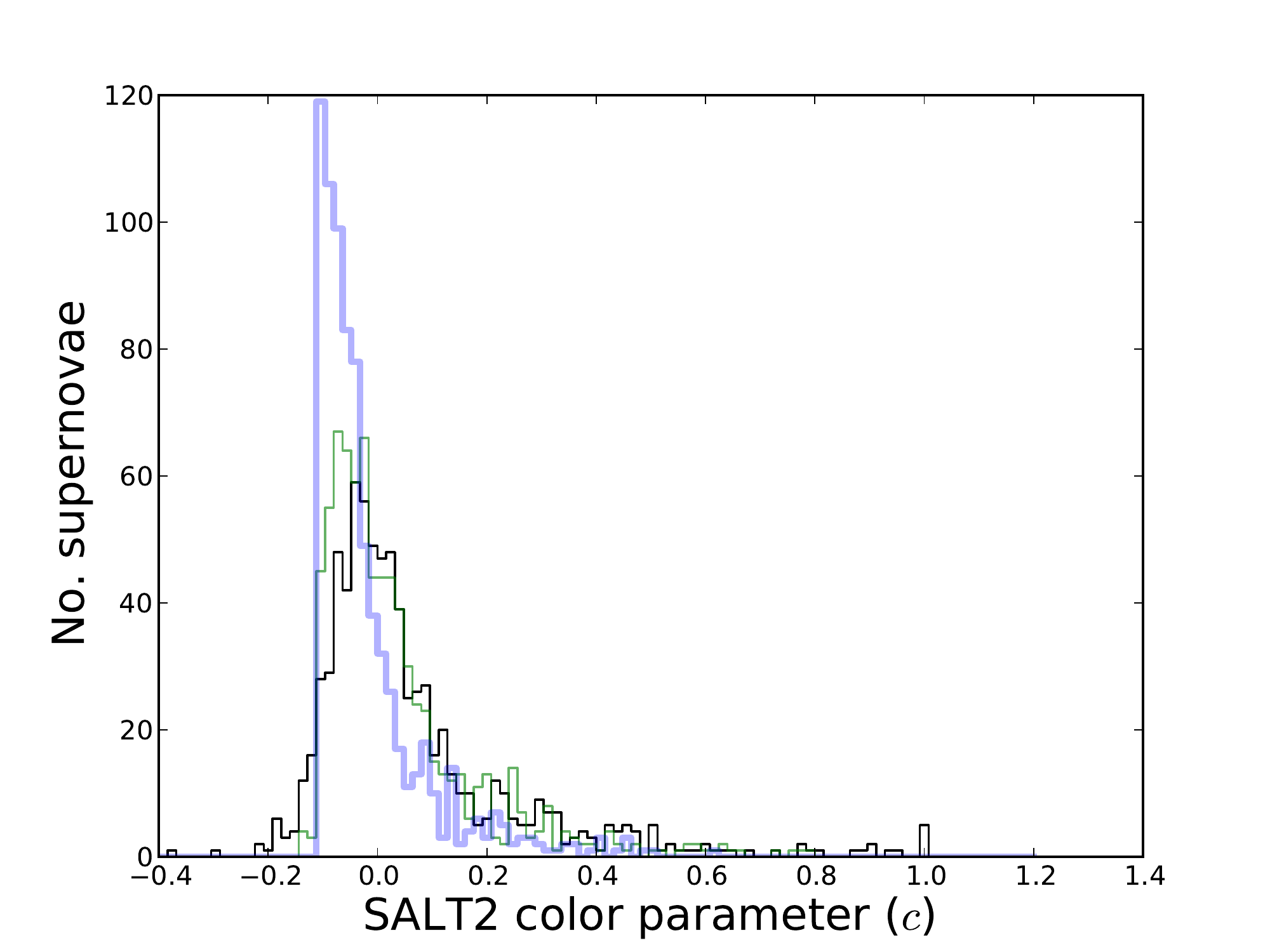}
\caption{The distribution of {\sc SALT2} color parameter c for all SN\,Ia (black line) and after converting the $A_V$ values from {\sc MLCS2k2} using equation \ref{eq:cA} (green histogram). Since the fit is reasonable, there is no big difference between the two. After correction for inclination alone, the $A_V$ values would translate into the blue histogram.}
\label{f:salt:histc}
\end{center}
\end{figure}

{\sc SALT2} handles the SN\,Ia color and dust extinction in a combined parameter, and it may not be obvious how our approach  has any easy application to this package. To illustrate, Figure \ref{f:salt:c} shows the relation between the observed $A_V$ and the {\sc SALT2} color parameter, c \citep[similar to Figure 9 in ][]{Sako08}. There is a linear relation between the two that follows:
\begin{equation}
\label{eq:cA}
c = 0.046*A_V + 0.012
\end{equation}
Thus, a change in the prior for $A_V$ in the case of {\sc MLCS2k2} would translate a change in the prior for {\sc SALT2}'s assumption but in color-space. To illustrate in figure \ref{f:salt:histc}, we plot the distribution of c values found in SDSS-III for SN\,Ia, as well as the values of $A_V$ from {\sc MLCS2k2} translated using the above relation. The two distributions are --unsurprisingly-- very similar. however, if we then translate the $A_V$ value corrected for the inclination of the disk (inferred from the axis ratio), the distribution of c values changes significantly. In our view, this change in distribution after inclination correction is the original distribution one could adopt as the prior distribution of the parameter $c$ for SALT2 applications, similar to the one to be adopted for {\sc MLCS2k2} (since both of them are essentially color priors).
%
%
%

\section{\label{s:discussion}Discussion}

The SN\,Ia community is well aware of the effects of host galaxy extinction may have on the results and have actively been searching for both ``dust-free" supernovae \citep[i.e., hosted by early-type galaxies, see e.g.,][]{Suzuki12}  and pushed light-curve observations into the near-infrared to mitigate extinction issues.

This paper's aim is an attempt to map out how the available observational information, and specifically the host galaxy information, can be further used to optimize the SN\,Ia distance measurement. As a first exercise, we applied the three available templates from occulting galaxies, naively matched to the SN\,Ia host through the stellar mass or galactocenric radius. Introducing new host extinction templates certainly has an effect but since the scatter in the distances is not markedly reduced, using just these templates would just skew the Cosmological result, not improve it. Despite the fact that the $A_V$ values were attained using a more Milky-Way type relation \citep{Kessler09}, we cannot rule out that none of the current scatter comes from host attenuation and all of it comes from intrinsic color scatter in the SN\,Ia.
But assuming much of the scatter is due to host attenuation, when matching host templates through projected radius rather than host mass, the scatter increases more than through mass-matching. This points, in our opinion, in which order one should match the occulting galaxy templates; first through mass and then through galactocentric radius.

Another key issue for SN\,Ia cosmological measurements will be if the extinction prior will change with redshift. Given that between z=0 and z=1, specific star-formation rates increased by a factor 3-10 \citep[e.g.,][]{Noeske07b}, one can expect a much more fractured dusty ISM \citep[thanks to increased turbulence, as we speculated in][]{Holwerda08a} or a much denser one (to sustain the star-formation). Some initial results on dust lanes in L$^*$ edge-ons suggest that the global morphology in disk galaxies remains the same: 80\% of these have a dust lane mid-plane \citep{Holwerda12b}. 

Alternatively, if the $A_V$ distribution is significantly different for massive versus lower-mass disk galaxies, any evolution in the mix of SN\,Ia host galaxies will result in an effective change in the appropriate $A_V$ prior. Dust in is strongly suspected to be distributed differently with respect to the stars in lower-mass galaxies. Initial Spitzer observations showed very extended dust morphology in a few individual cases of lower-mass galaxies \citep{Hinz07,Hinz09}, similar to the occulting pair in Figure \ref{f:pair}. Recently, it has become clear that in the stacked SED of lower-mass galaxies there is virtually no attenuation in the optical and ultra-violet yet Herschel emission clearly show large dust masses at very low temperatures \citep[][]{Popescu02b, Grossi10,Bourne12a}. The combined evidence points to a dependence of star/dust geometry on galaxy mass. Whether this means dust is relatively more extended in radius, height above the plane \citep[e.g.,][]{Dalcanton04,Holwerda12a} or clumpiness, remains to be seen.

Fortunately, global extinction measures can be checked using high-redshift occulting galaxy pairs and both effects can then be accounted for, be it evolution in the mix of host galaxies or in the ISM of the host galaxies themselves. There is, at presence, no observed evolution in the color distribution of observed SN\,Ia \citep[][]{Conley11,Betoule14}. However, {\em if} there is significant evolution in the $P(A_V)$ distribution with redshifts --either due to ISM evolution in the host galaxies or a significant change in the host galaxy population mix--, it may change the relative luminosity distances as a function of redshift, the driver or dark energy cosmology.

We hope to have convinced the reader that the use of occulting galaxy pairs to obtain extinction distribution templates is a good way forward to help reduce the remaining uncertainties in SN\,Ia distance measurements. The eventual goal is to cover enough range in stellar mass, radial coverage and inclinations to map the mean and variance of the typical extinction distribution in spiral galaxy disks. This will have many astrophysical applications but the use as a prior for Cosmological SN\,Ia measurements was the principal driver for us to obtain HST observations of occulting pairs: GO-13695, {\em STarlight Absorption Reduction through a Survey of Multiple Occulting Galaxies (STARSMOG)} P.I. B.W. Holwerda, a 150 orbit SNAPshot program \citep{Holwerda14b}, which is projected to  complete a sample of 98 occulting pairs by 2017.

\section*{Acknowledgements}

The authors are very grateful for the insightful comments of the anonymous referee and to the SDSS-SN team who provided the data.
KJM is supported by an Australian Research Council Discovery Early Career Researcher Award.
This research has made use of the NASA/IPAC Extragalactic Database (NED) which is operated by the Jet Propulsion Laboratory, California Institute of Technology, under contract with the National Aeronautics and Space Administration. 
This research has made use of NASA's Astrophysics Data System.
This research made use of Astropy, a community-developed core Python package for Astronomy \citep{Astropy-Collaboration13a}. This research made use of matplotlib, a Python library for publication quality graphics \citep{Hunter07}. PyRAF is a product of the Space Telescope Science Institute, which is operated by AURA for NASA. This research made use of SciPy \citep{scipy}. 

%

\end{document}